# Correlated electron dynamics with time-dependent quantum Monte Carlo: three-dimensional helium

Ivan P. Christov

Physics Department, Sofia University, 1164 Sofia, Bulgaria

**Abstract**

Here the recently proposed time-dependent quantum Monte Carlo method is applied to three dimensional para- and ortho-helium atoms subjected to an external electromagnetic field with amplitude sufficient to cause significant ionization. By solving concurrently sets of up to 20 000 coupled 3D time-dependent Schrödinger equations for the guide waves and corresponding sets of first order equations of motion for the Monte Carlo walkers we obtain ground state energies in close agreement with the exact values. The combined use of spherical coordinates and B-splines along the radial coordinate proves to be especially accurate and efficient for such calculations. Our results for the dipole response and the ionization of an atom with un-correlated electrons are in good agreement with the predictions of the conventional time-dependent Hartree-Fock method while the calculations with correlated electrons show enhanced ionization that is due to the electron-electron repulsion.

# 1. Introduction

The development of time-dependent methods in physics and chemistry is of primary interest in areas where the non-perturbative electron dynamics in atoms, molecules, and semiconductors is important. These areas include atoms and molecules driven by laser fields, quantum transport through molecular junctions, and nano-scaled electronic devices. That research is further inspired by the recent progress in femto and attosecond lasers that made it possible to explore time-dependent phenomena with unprecedented temporal resolution of tens of attoseconds, which has the potential to capture the motion of electrons exposed to Coulomb and exchange forces due to other electrons and nuclei. When solving the many-body time dependent Schrödinger equation (TDSE) it is presumed, according to the statistical interpretation of quantum mechanics, that we seek for the evolving coordinates of an infinite ensemble of electrons which represent each separate electron degree where the statistical distribution of these particles is given by modulus square of the wave function. Since the many-body wave function $\Psi(\mathbf{R},t)$ resides in configuration space with arguments being the instantaneous coordinates of all electrons $\mathbf{R} = (\mathbf{r}_1, \mathbf{r}_2,...,\mathbf{r}_N)$, the computational workload for solving the corresponding TDSE scales exponentially with the number of electrons, typically as $K^{3N}$ for N particles on a grid of K spatial nodes. This scaling is usually explained to be a result of specific quantum non-locality effects which take place in the many-body system. Since the exponential scaling poses insurmountable difficulties for finding accurate numerical solution for more than two particles in three spatial dimensions approximate methods have been introduced which scale polynomially with time. These methods include time-dependent Hartree-Fock (TDHF) [1] and time-dependent density functional theory (TDDFT) [2] where the many-body problem is reduced to single-body problems of non-interacting electrons moving in averaged potential for TDHF, or in effective but generally unknown exchange-correlation potential for TDDFT. Other methods use series expansions over multiple configurations in order to account for the time-dependent correlation effects [3,4] which, however, requires the calculation of large number of Coulomb and exchange integrals that lowers their efficiency dramatically.

The quantum Monte Carlo (QMC) methods employ large but finite number of particles (walkers) to calculate the probability distributions of many-body quantum states [5]. In diffusion quantum Monte Carlo (DMC) the walkers relax towards the ground state due to the combination of diffusion and branching events in which the number of walkers at a given point is proportional to $|\Psi(\mathbf{R},\tau)|$. With the introduction of an auxiliary guiding function $\Psi_G(\mathbf{R})$ it can be shown that the particle distribution is given by the product $f(\mathbf{R},\tau) = \Psi_G(\mathbf{R})\Psi(\mathbf{R},\tau)$ which may approach the statistical distribution in configuration space $|\Psi(\mathbf{R},\tau)|^2$, for appropriately chosen function $\Psi_G(\mathbf{R})$. However, this strategy cannot be used for real-time propagation. Recently, new method to solve many-body quantum problems *ab initio* which uses simultaneously evolving particles and guide waves was introduced [6-8]. In this method the many-body TDSE is reduced to a set of coupled single-body TDSE for the guide waves $\varphi^k(\mathbf{r},t)$ and equations of motion for the Monte Carlo (MC) particles with trajectories $\mathbf{r}^k(t)$, where each particle (walker) is attached to a separate guiding wave. This approach allows both imaginary and real-time evolution of the waves together with the motion of the walkers to be calculated self-consistently, with both local and non-local quantum correlation effects fully accounted for. In fact, the time dependent quantum Monte Carlo (TDQMC) method recovers the symmetry that is due to the particle-wave dualism in quantum mechanics where the calculation is performed in physical space for both particles and associated guiding waves. It is assumed in TDQMC that each walker samples its own distribution given by the modulus square of the corresponding guiding wave $|\varphi^k(\mathbf{r},t)|^2$ that obeys 3D time dependent Schrödinger equation. The many-body probability distribution in configuration space is then considered to be an intersection of these mutually connected single-body distributions, each one represented by the corresponding Monte Carlo particle with trajectory $\mathbf{r}^k(t)$. After a multi-variate kernel density estimation (KDE) over the discrete distribution of particles a continuous distribution function is obtained which approaches the module square of the many-body quantum state. This method can be considered to be an extension of the statistical interpretation of quantum mechanics to the case of coupled distributions which are described by a set of coupled Schrödinger equations. It is important to stress that since in TDQMC the Monte Carlo walkers are guided by first-order de Broglie-Bohm equations (that do not involve quantum potentials) its predictions need not be related to the Bohmian mechanics and its interpretations [9,10]. Important advantage of TDQMC is that it allows treatment of complex quantum-classical systems of different kinds of particles (e.g. electrons and nuclei) without invoking the Born-Oppenheimer approximation [11]. In this paper we apply the TDQMC method to three dimensional atoms in external field by using spherical coordinates, that greatly improves the efficiency of the calculations. Para- and ortho- helium atoms are considered as an example where the results are compared with the predictions of the time-dependent

Hartree-Fock method. Other work that consideres 3D helium atom in strong external field includes time-dependent close coupling method [12] and expansion over coherent states [13].

## 2. General theory

Here we consider correlated electron dynamics in a three dimensional multi-electron atom subjected to an external electromagnetic field, although the same method can be applied to molecules and other quantum objects. Within the fixed-nuclei approximation, the $N$-electron system is described by the many-body time dependent Schrödinger equation:

$$i\hbar \frac{\partial}{\partial t}\Psi(\mathbf{R},t) = -\frac{\hbar^2}{2m}\nabla^2\Psi(\mathbf{R},t) + V(\mathbf{R})\Psi(\mathbf{R},t) \ , \qquad (1)$$

where $\mathbf{R} = (\mathbf{r}_1,...,\mathbf{r}_N)$ is a $3N$ dimensional vector in configuration space which specifies the coordinates of $N$ electrons, and $\nabla = (\nabla_1, \nabla_2,..., \nabla_N)$. The Hamiltonian in Eq. (1) is inseparable in the electron coordinates, due to the electron-electron interaction:

$$V(\mathbf{r}_1,...,\mathbf{r}_N) = V_{e-n}(\mathbf{r}_1,...,\mathbf{r}_N) + V_{e-e}(\mathbf{r}_1,...,\mathbf{r}_N) + V_{ext}(\mathbf{r}_1,...,\mathbf{r}_N,t)$$

$$= \sum_{k=1}^{N} V_{e-n}(\mathbf{r}_k) + \sum_{k>l}^{N} V_{e-e}(\mathbf{r}_k - \mathbf{r}_l) + V_{ext}(\mathbf{r}_1,...,\mathbf{r}_N,t) \ , \qquad (2)$$

where the many-body potential in Eq. (2) is a sum of electron-nuclear, electron-electron, and external potentials. The TDQMC approach to many-electon atoms assigns ensembles of classical walkers to each electron where the k-th walker from the i-th electron ensemble follows a definite trajectory $\mathbf{r}_i^k(t)$ which samples its own statistical distribution given by the modulus square of the corresponding guiding wave $\left|\varphi_i^k(\mathbf{r},t)\right|^2$. On the other side, the guiding waves obey a set of coupled time dependent Schrödinger equations in physical space where the walker's trajectories participate in the electron-electron potential. Thus, unlike in other methods, TDQMC uses configurations of particles and guiding waves in a symmetric manner without invoking explicit series expansions over multiple configurations that would require expensive calculation of various volume integrals. Each replica of the many-body quantum state can be represented as an anti-symmetrised product of the corresponding guide waves:

$$\Psi^k(\mathbf{r}_1,\mathbf{r}_2,...,\mathbf{r}_N,t) = A\prod_{i=1}^{N}\varphi_i^k(\mathbf{r}_i,t), \qquad (3)$$

which determines the velocity of the k-th walker form i-th electron ensemble through the first-order de Broglie-Bohm guidance equation:

$$\dot{\mathbf{r}}_i^k(t) = \frac{\hbar}{m}\mathrm{Im}\left[\frac{1}{\Psi^k(\mathbf{r}_1,...\mathbf{r}_j...,\mathbf{r}_N,t)}\nabla_i\Psi^k(\mathbf{r}_1,...\mathbf{r}_j...,\mathbf{r}_N,t)\right]_{\mathbf{r}_j=\mathbf{r}_j^k(t)}, \qquad (4)$$

where i,j=1,…,N denotes the electron, and k=1,…,M denotes the Monte Carlo walker under consideration. The separate guide waves obey a set of coupled TDSE [8]:

$$i\hbar \frac{\partial}{\partial t} \varphi_i^k(\mathbf{r}_i,t) = \left[ -\frac{\hbar^2}{2m} \nabla_i^2 + V_{e-n}(\mathbf{r}_i) + \sum_{j \neq i}^{N} V_{e-e}^{eff}[\mathbf{r}_i - \mathbf{r}_j^k(t)] + V_{ext}(\mathbf{r}_i,t) \right] \varphi_i^k(\mathbf{r}_i,t), \qquad (5)$$

where the effective electron-electron potential in Eq. (5) connects the separate TDSE that describe different electrons. Since each electron is represented by a large but finite number of M walkers, each of these walkers should feel the potential due to a number of $M_1 \leq M$ walkers which belong to another electron, which in fact translates the quantum non-locality effects to the language of coupled particle ensembles. The effective electron-electron potential in Eq. (5) can then be expressed as a Monte Carlo sum over the smoothed distribution of the Coulomb potentials [8]:

$$V_{e-e}^{eff}[\mathbf{r}_i - \mathbf{r}_j^k(t)] = \frac{1}{Z_j^k} \sum_{l=1}^{M_1} V_{e-e}[\mathbf{r}_i - \mathbf{r}_j^l(t)] \, K\!\left( \frac{\left|\mathbf{r}_j^l(t) - \mathbf{r}_j^k(t)\right|}{\sigma_j^k\!\left(\mathbf{r}_j^k,t\right)} \right), \qquad (6)$$

where the statistical weighting factor $Z_j^k$ is given by :

$$Z_j^k = \sum_{l=1}^{M_1} K\!\left( \frac{\left|\mathbf{r}_j^l(t) - \mathbf{r}_j^k(t)\right|}{\sigma_j^k\!\left(\mathbf{r}_j^k,t\right)} \right), \qquad (7)$$

where $K$ is a smoothing kernel which transforms discrete distributions (walker trajectories) to continuous distributions (functions). Clearly, the widths $\sigma_j^k\!\left(\mathbf{r}_j^k,t\right)$ of the kernel in Eq. (6) and Eq. (7) are measures for the length of nonlocal quantum correlations between the ensembles of walkers which represent different electrons. In this way the quantum non-locality is manifested as a Coulomb interaction of the k-th walker from the j-th electron ensemble not only with the k-th walkers from the ensembles that represent the rest of the electrons, but also with other walkers from these ensembles that lie within the range of the non-local correlation length $\sigma_j^k\left(\mathbf{r}_j^k,t\right)$. As a result, trajectory entanglement occurs that stabilizes the ground state of the quantum system. In fact, the smoothing kernels in Eq. (6) transform the Coulomb potential due to discrete walker distribution to an effective potential of a continuous charge distribution (infinite number of walkers), where the length $\sigma_j^k\left(\mathbf{r}_j^k,t\right)$ can be determined by a kernel density estimation over the Monte Carlo data. When $\sigma_j^k\left(\mathbf{r}_j^k,t\right)\to\infty$ the effective potential of Eq. (6) is reduced to the Hartree potential given by a simple sum of the Coulomb potentials due to the MC walkers.

Thus in the TDQMC method many replicas of the trial wave function constructed by the guide waves (different k in Eq.(3)) are generated, and one walker is picked up which belongs to the probability distribution given by each separate guide wave. The set of these walkers represents the joined probability density of the distributions given by the guide waves, which corresponds to the correlated probability density of the many-body quantum system. Since the different replicas of the wave function $\varphi_i^k(\mathbf{r},t)$ in Eq. (5) move in different potentials these are generally not orthogonal for the different electron states. However, each walker possesses its own set of ortho-normal guide waves where for the k-th walker we have:

$$\int \varphi_i^{k*}(\mathbf{r},t)\varphi_j^k(\mathbf{r},t)d\mathbf{r}=\delta_{ij}, \tag{8}$$

which is important for the calculation of parallel-spin electron dynamics. Besides using anti-symmetrized states (Eq. (3)) one can incorporate the fermionic effects by employing screened potentials where the electron-electron repulsion is reduced due to the exchange non-locality [14]. In this case the size of the exchange hole around each walker can be introduced explicitly, usually starting from the Hartree-Fock approximation, which can be helpful for the interpretation of various many-body quantum effects.

## 3. Method of solution

Since for practical applications the TDQMC method involves simultaneous solution of large number of time-dependent Schrödinger equations for the guide waves, efficient numerical methods are needed, particularly for three spatial dimensions. Here we present such a method which offers very good accuracy, which can be applied for both atomic and molecular systems.

### A. Real-space calculation in spherical coordinates

Since the atomic and many molecular systems are most naturally described in spherical coordinates with independent variables $\mathbf{r} = (r, \theta, \varphi)$, we use standard single-center expansion for each of the time-dependent guide waves of Eq. (5):

$$\varphi(\mathbf{r},t) = \sum_{l=0}^{\infty} \sum_{m=-l}^{l} \frac{R_l^m(r,t)}{r} Y_l^m(\theta,\varphi) \qquad (9)$$

where $Y_l^m(\theta,\varphi)$ are spherical harmonics depending on the angular coordinates. The radial function $R_l^m(r,t)$ is then a solution of the reduced TDSE (atomic units $e = m = \hbar = 1$ are used henceforth):

$$i\frac{\partial}{\partial t} R_l^m(r,t) = \left[ -\frac{1}{2}\frac{d^2}{dr^2} - \frac{Z}{r} + \frac{l(l+1)}{2r^2} \right] R_l^m(r,t) + \sum_{l',m'} V_{l,l'}^{m,m'}(r,t) R_{l'}^{m'}(r,t), \qquad (10)$$

where Z denotes the charge of the nucleus and $V_{l,l'}^{m,m'}$ are the matrix elements of the total potential:

$$V(\mathbf{r},t) = \sum_{i=1}^{N} V_{e-e}^{eff}[\mathbf{r} - \mathbf{r}_i(t)] + V_{ext}(\mathbf{r},t). \qquad (11)$$

For Coulomb interaction where $V_{e-e}[\mathbf{r} - \mathbf{r}_i(t)] = |\mathbf{r} - \mathbf{r}_i(t)|^{-1}$ we have in Eq. (10):

$$V_{l,l'}^{m,m'}(r,t) = \int_0^{\pi} d\theta \int_0^{2\pi} d\varphi\, Y_l^{m*}(\theta,\varphi) Y_{l'}^{m'}(\theta,\varphi) \sin(\theta) \left[ \sum_{i=1}^{N} \frac{1}{|\mathbf{r} - \mathbf{r}_i(t)|} + V_{ext}(\mathbf{r},t) \right], \qquad (12)$$

where, in general, the angular integrals can be reduced to series over products of 3j-symbols by using the multipole expansion of the electron-electron Coulomb repulsion [15]:

$$\frac{1}{|\mathbf{r}-\mathbf{r}_i(t)|} = \sum_{l=0}^{\infty}\sum_{m=-l}^{l} \frac{4\pi}{2l+1} \frac{r_<^l}{r_>^{l+1}} Y_l^m(\theta,\varphi) Y_l^{m*}(\theta_i,\varphi_i), \qquad (13)$$

where $r_< = \min[r, r_i(t)]$, $r_> = \max[r, r_i(t)]$, and $(\theta_i, \varphi_i)$ are the angular coordinates of the particle at $\mathbf{r}_i(t)$. However, it is seen from Eq. (12) that in general the potential matrix $V_{l,l'}^{m,m'}(r,t)$ is not diagonal with respect to the quantum number m which is not preserved, which reflects the lack of azimuthal symmetry for arbitrary trajectory $\mathbf{r}_i(t)$ of the walker. This so called "m-mixing" problem can be resolved in our case by rotating the coordinate system so that in the rotated coordinates the particle lies on the z axis $\mathbf{r}_i(t) \to \mathbf{r}'_i(t) = (0,0,z')$. This transformation maps the generally asymmetric potential into such with azimuthal (cylindrical) symmetry where the quantum number m is preserved (see also [16]) and the matrix $V_{l,l'}^{m,m'}(r,t)$ becomes diagonal with respect to m. Then the equations for the radial functions can be separated for the different m:

$$i\frac{\partial}{\partial t}R_l^m(r,t) = \left[-\frac{1}{2}\frac{d^2}{dr^2} - \frac{Z}{r} + \frac{l(l+1)}{2r^2}\right] R_l^m(r,t) + \sum_{l'} V_{l,l'}^m(r,t) R_{l'}^m(r,t), \qquad (14)$$

where:

$$V_{l,l'}^m(r,t) = (-1)^m \sqrt{(2l+1)(2l'+1)} \sum_{l''} \frac{r_<^{l''}}{r_>^{l''+1}} \begin{pmatrix} l & l' & l'' \\ 0 & 0 & 0 \end{pmatrix} \begin{pmatrix} l & l' & l'' \\ -m & m & 0 \end{pmatrix}$$

$$+ D_{l,l'}^m r E(t), \qquad (15)$$

where the parentheses denote the Wigner 3j symbols and $D_{l,l'}^m rE(t)$ is the external potential in the dipole approximation with $E(t)$ being the external electric field. The matrix of the dipole moment $D_{l,l'}^m$ has non-zero upper and lower blocks in the spherical basis:

$$D_{l,l+1}^m = \sqrt{\frac{(l+1)^2 - m^2}{(2l+1)(2l+3)}}; \qquad l=0,1,\ldots \qquad (16)$$

$$D_{l,l-1}^m = \sqrt{\frac{l^2 - m^2}{(2l-1)(2l+1)}}; \qquad l=1,2,\ldots \qquad (17)$$

**B. Introducing B-spline basis functions**

In order to attain higher accuracy when solving Eq.(14) for singular potentials we expand the radial functions $R_l^m(r,t)$ onto the B-spline basis set [17]:

$$R_l^m(r,t) = \sum_i c_i^{lm}(t) B_i(r), \qquad (18)$$

where $B_i(r)$ is a B-spline. After substituting Eq. (18) into Eq. (14) and integrating over the radial coordinate, we obtain (in matrix notations):

$$i\mathbf{S}\frac{\partial \mathbf{c}^{lm}}{\partial t} = \mathbf{H}_0^l \mathbf{c}^{lm} + \sum_{l'} \mathbf{V}_{l,l'}^m \mathbf{c}^{l'm}, \qquad (19)$$

where the matrix elements in the spline basis are given by:

$$(\mathbf{S})_{ij} = \int_0^{r_{max}} B_i(r) B_j(r) dr \tag{20}$$

$$(\mathbf{H}_0)_{ij} = -\frac{1}{2} \int_0^{r_{max}} B_i(r) \frac{d^2}{dr^2} B_j(r) dr + \int_0^{r_{max}} \left[ -\frac{Z}{r} + \frac{l(l+1)}{2r^2} \right] B_i(r) B_j(r) dr; \tag{21}$$

$$(\mathbf{V})_{ij} = (-1)^m \sqrt{(2l+1)(2l'+1)} \sum_{l''} \begin{pmatrix} l & l' & l'' \\ 0 & 0 & 0 \end{pmatrix} \begin{pmatrix} l & l' & l'' \\ -m & m & 0 \end{pmatrix} \int_0^{r_{max}} B_i(r) \frac{r_<^{l''}}{r_>^{l''+1}} B_j(r) dr$$

$$+ (\mathbf{D})_{ij} E(t) \int_0^{r_{max}} B_i(r) r B_j(r) dr, \tag{22}$$

where all matrix elements can be calculated exactly using Gaussian quadrature. It should be noted that due to the compact support of the B-spline basis all matrices in Eq. (19) are symmetric and have block structure that is very favorable for both numerical storage and computations. Another advantage of using splines is that the gradients in Eq.(4) are calculated with the machine accuracy.

## 4. Results and discussion

In order to observe the time evolution of three-dimensional para- and ortho-helium atoms in external electromagnetic field we solve numerically the reduced Schrödinger equation for the guide waves (Eq. (19)) concurrently with the guiding equation for the Monte Carlo walkers (Eq. (4)). Depending on the symmetry of the ground state, we have for each replica of the two-body spatial wave function:

$$\Psi^k(\mathbf{r}_1,\mathbf{r}_2,t) = \frac{1}{\sqrt{2}}\left[\varphi_1^k(r_1,l,m,t)\varphi_2^k(r_2,l,m,t) \pm \varphi_1^k(r_2,l,m,t)\varphi_2^k(r_1,l,m,t)\right], \qquad (23)$$

where the plus sign is for a symmetric spin-singlet ground state (para-helium) and the minus is for parallel-spin (ortho-helium) ground state.

Since the algorithm used here involves rotation of the wave function, we reduce the time propagation to a sequence of time sub-steps (time split-stepping):

$$\mathbf{c}^{lm}(t+\Delta t) = e^{-i\mathbf{H}_0\Delta t/2} e^{-i\mathbf{V}(t+\Delta t/2)\Delta t} e^{-i\mathbf{H}_0\Delta t/2} \mathbf{c}^{lm}(t), \qquad (24)$$

where the exponentiation in Eq. (24) is performed efficiently using a second-order in time implicit linear system solver based on the biconjugate gradient stabilized method with preconditioning (BiCGSTAB [18]) which benefits from the sparseness of the matrices. The coordinate rotation needed to apply the potential term in Eq. (24) is accomplished by multiplying the radial function $R_l^m(r,t)$ of Eq. (18) by the Wigner rotation matrices $W_l^m(\theta_i,\varphi_i)$ built from the angular coordinates of the corresponding walker $(\theta_i,\varphi_i)$, that in fact rotates the whole wave function $\varphi(\mathbf{r}_i,t)$ synchronously with the particle rotation.

After the potential due to each walker is applied the wave function is returned back to its original position by a backward rotation. It should be noted that these rotation are numerically very fast and their total number equals the number $M_1$ in Eq. (6). One additional rotation is required for applying the linearly polarized external field.

First, the walker's distribution for the ground state of the atom is calculated by choosing an initial set of guide waves $\varphi_i^k(\mathbf{r}_i, t=0)$ close to the eigen-functions of the Hamiltonian $H_0$ for the 1s and 2s states, and then preparing the initial ensembles of Monte Carlo walkers for each electron $\mathbf{r}_i^k(t=0)$. Next the system of particles and waves is propagated in complex time until steady state in electron energy is established. The use of complex time $t = t' + it''$ during the ground state preparation ensures that the guide waves relax to the ground state owning to $t''$ while each of these waves acquires a time-dependent phase due to $t'$ which guides the walkers to their stationary positions through Eq. (4) (where the time variable is also $t'$). A random component in added to the walker's motion that thermalizes the ensemble to avoid possible bias in the walker distribution that may arise due to the quantum drift alone. In practice Metropolis algorithm is used to sample the densities $\left|\varphi_i^k(\mathbf{r}_i, t)\right|^2$ at each time step. Best result for the ground state distribution is expected when the drift and the diffusion are in balance which is achieved for $t' = t''$. Since at steady state both the drift and the diffusion velocities of the walkers should vanish, the amplitude of the random component fades out with time. Once steady state is established, the imaginary time component $t''$ is set to zero, and the evolution of the system proceeds in real time for both guide waves and particles. As noted before the kernel density estimation is an essential part of the quantum calculations in that it smoothes the walker's distributions and determines the characteristic dimensions of non-local coupling between the walkers from ensembles that belong to different electrons. The width $\sigma_j^k\left(\mathbf{r}_j^k,t\right)$ in Eq. (6) and Eq. (7) depends on the electron density (the density of walkers) in the quantum system. For Gaussian kernels $\sigma_j^k\left(\mathbf{r}_j^k,t\right)$ can be estimated using a simple formula [19]:

$$\sigma_j^k(\mathbf{r},t) = \sigma \sqrt{\frac{g_j}{\rho_j^k(\mathbf{r},t)}}, \qquad (25)$$

where $\rho_j^k(\mathbf{r},t)$ is a pilot density estimate of the walker distributions for the $j$-th electron, which can be obtained using kernel density estimation with constant bandwidth $\sigma$ which depends on the statistical properties of the walker's ensemble, and $g_j$ are the geometric means of the values of $\rho_j^k(\mathbf{r},t)$, for $k=1,...,M$, respectively. The widths $\sigma_j^k\left(\mathbf{r}_j^k,t\right)$ can be calculated either by applying separate one-dimensional KDE's along each spatial direction or by using the covariance matrices for the corresponding walkers. Since the walker's distribution is what determines the correlated electron density that results from the electron-nuclear attraction and electron-electron repulsion, the widths $\sigma_j^k\left(\mathbf{r}_j^k,\tau\right)$ at moment $\tau$ where steady state is established can be used to estimate the energy of the ground state without referencing to the guide waves. We start with the product kernel estimator of the following form:

$$P(\mathbf{r}_1,\mathbf{r}_2,...,\mathbf{r}_N,\tau) = \frac{1}{M}\sum_{k=1}^{M}\left\{\prod_{i=1}^{N}\prod_{d=1}^{D}\frac{1}{\sqrt{2\pi}\sigma_{i,d}^{k}\left(\mathbf{r}_{i,d}^{k},\tau\right)}\exp\left[-\frac{\left|\mathbf{r}_{i,d}(\tau)-\mathbf{r}_{i,d}^{k}(\tau)\right|^2}{2\sigma_{i,d}^{k}\left(\mathbf{r}_{i,d}^{k},\tau\right)^2}\right]\right\}, \qquad (26)$$

where the index d=1,2…D denotes the axes in physical space of dimension D. Equation (26) describes the many-body particle density $P(\mathbf{R},\tau)$ of M walkers in the configuration space, which after substitution into the expression for the energy:

$$E(\tau) = \int P(\mathbf{R},\tau)\left[\frac{1}{8}\frac{[\nabla_\mathbf{R} P(\mathbf{R},\tau)]^2}{P(\mathbf{R},\tau)^2} + V(\mathbf{R},\tau)\right]d\mathbf{R}, \qquad (27)$$

reduces it to a MC sum over the steady state walker's positions:

$$E(\tau) = \frac{1}{M}\sum_{k=1}^{M}\left[\frac{1}{8}\sum_{i=1}^{N}\frac{[\nabla_{\mathbf{r}_i} P(\mathbf{r}_1,\mathbf{r}_2,...,\mathbf{r}_N,\tau)]^2}{P(\mathbf{r}_1,\mathbf{r}_2,...,\mathbf{r}_N,\tau)^2}\bigg|_{\mathbf{r}_i=\mathbf{r}_i^k(\tau)} \right.$$

$$\left. +\sum_{i=1}^{N}V_{e-n}(\mathbf{r}_i^k)\bigg|_{\mathbf{r}_i^k=\mathbf{r}_i^k(\tau)} + \sum_{i>j}^{N}V_{e-e}(\mathbf{r}_i^k-\mathbf{r}_j^k)\bigg|_{\mathbf{r}_{i,j}^k=\mathbf{r}_{i,j}^k(\tau)}\right] \qquad (28)$$

In fact, the major advantage of using KDE is that the gradients in Eq.(28) are calculated analytically through differentiation of the kernel function (often Gaussian as in Eq. (26)), without invoking finite differences of multi-variate functions that would introduce larger numerical error.

We calculate the symmetric ground state of 3D para-helium atom by propagating the initial ensembles of walkers and guide waves in complex time as described before. A spatial grid size of 20 a.u. for para-helium and 30 a.u. for ortho-helium, and a complex-time step size (0.05 au, –0.05 au) are used where the radial coordinate is covered by up to 100 cubic splines. The radial grid ends with absorbing boundaries for both the waves and the particles. Figure 1 shows with blue lines the radial distribution density (a smoothed histogram) of M=20 000 Monte Carlo particles after 400 time steps where steady state is established for the time-dependent Hartree-Fock approximation ($\sigma_j^k(\mathbf{r}_j^k,t) \to \infty$ in Eq. (6) and Eq. (7)) and for the correlated TDQMC calculation (red lines). It is seen from Fig. 1 (a), (b) that for both para- and ortho-helium the TDQMC distributions are slightly broader as compared to the TDHF. That broadening can be easily explained to be a result of the correlated electron motion where the Coulomb repulsion pushes the particles away from each other while positioning them at the opposite sides with respect to the core, which effectively increases the distance between the electrons as compared to their distance to the core, and also reduces the shielding of the nuclear charge. As a result the electron-nuclear attraction increases, which increases the density of the electrons around the nucleus and so increases the binding energy. These effects are more pronounced for ortho-helium (Fig. 1 (b)) where the outer electron is more weakly bound and there is also a contribution due to the repulsive exchange forces. We found -2.91 a.u. (79.15 eV) for the energy of the correlated ground state of para-helium to be compared with the Hartree-Fock result of -2.86 a.u. . For ortho-helium ground state energy of -2.2 a.u. (59.9 eV) versus -2.17 a.u. for Hartree-Fock were obtained. It should be noted that for 20 000 walkers the calculated energies fluctuate within 0.05 a.u. between the different runs so that the above values of the energy have been averaged over 10 runs.

Next we show in Figure 2 the results for the time-dependent ionization and dipole moment of para- and ortho-helium obtained from the real-time solution of the reduced Schrödinger equations (Eq. (19)) together with the guiding equations (Eq. (4)). We use linearly polarized electromagnetic field $E(t) = E_0 \sin(\omega t)$ with carrier frequency 0.153 a.u. and duration of two periods (see the inset of Fig. 2 (a)). Figure 2 (a) shows the survival probability as function of time for para-helium exposed to electromagnetic field with peak amplitude $E_0$=0.4 a.u., that is sufficient to cause significant tunneling ionization (above 10%). The survival probability is calculated as a portion of the MC walkers which remains within the numerical grid during the interaction. The green and red lines show the results from the conventional time-dependent Hartree-Fock (TDHF) and from the uncorrelated TDQMC ($\sigma_j^k\left(\mathbf{r}_j^k, t\right) \to \infty$), respectively. It is seen that the green and the red curves are very close for shorter times while for longer times the TDQMC calculation somewhat overestimates the ionization, that can be attributed to the reduced number of walkers $M_1$ in Eq. (6) and Eq. (7) (in this calculation we use $M_1$=30). The blue line in Fig. 2 (a) shows the correlated TDQMC result which predicts by 10% lower survival probability due to the electron correlation which enhances the ionization. Figure 2 (b) shows the time-dependent dipole moment of the para-helium atom, which is calculated directly from the particle distribution:

$$\mathbf{d}_i(t) \propto \sum_{k=1}^{M} \mathbf{r}_i^k(t) \tag{29}$$

It is seen from Fig. 2 (b) that in the three above cases the dipole moment fades out due to the ionizing MC walkers while the dipole amplitude for the correlated TDQMC becomes highest during first half-period. This is a signature for the electron-electron repulsion which increases the distances $\mathbf{r}_i^k(t)$ in Eq. (29) over their TDHF values. The time dependent behavior of the ionization and the dipole moment for 3D ortho-helium is shown in Fig. 3 for much lower amplitude of the external field $E_0=0.03$ a.u. This is so because the ionizing electron here is in 2s state which is much weaker bound than the 1s state. Again, the correlated TDQMC case shows enhanced ionization that is due to the electron-electron repulsion (see Fig. 3 (a)). However, it can be seen from Fig. 3 (b) that the ionizing trajectories do not enhance the dipole moment as for para-helium (Fig. 2 (b)) because only the central part of the 2s state is positioned in proximity of the nucleus where it overlaps with the electron cloud of the 1s state, while at the same time the outer trajectories of the 2s state contribute most to the atomic dipole moment. The above results prove that the tunneling ionization, that dominates for these field strengths, is correctly described by the TDQMC method.

## 5. Conclusions

Here the recently proposed time-dependent quantum Monte Carlo method is applied to three-dimensional para- and ortho-helium atoms. We present an approach to solve efficiently and accurately large systems of coupled time-dependent Schrödinger equations for the guiding waves together with first order equations of motion for the Monte Carlo walkers. The use of effective potentials allows us to account for the dynamic local and nonlocal quantum correlations between the electrons where kernel density estimation is used to calculate the corresponding nonlocal correlation widths. Our calculations predict correctly the correlated ground state energies, the time dependent ionization, and the time-dependent dipole moment of the atoms subjected to external electromagnetic field.

Since the probability density of the correlated many-body quantum state is determined by the spatial distribution of Monte Carlo walkers, other powerful particle-based methods such as the Ewald summation method can be used to ease the calculation of the Coulomb potentials in many-electron systems [20]. Also, the essential parallelism of the TDQMC method is found to be very helpful for systems where tens of thousands of MC walkers are needed in order to achieve sufficient accuracy. This is so because the TDQMC calculations require little communications between the parallel processes which solve different groups of TDSE's, mostly for calculation of the nonlocal quantum correlation effects. We have verified that by using massively parallel supercomputer (BlueGene/P) up to 20 000 coupled three dimensional time dependent Schrödinger equations can be solved simultaneously for affordable time with an excellent scalability. In this way the exponentially time-scaled quantum many-body problem that would require calculation of both the amplitude and the phase of the many-body quantum state is reduced to solving the problem of correlated probability density in physical space, which can be resolved with polynomial time-scaling, and is therefore accessible to a classical computer.

## Acknowledgments

The author gratefully acknowledges support from the National Science Fund of Bulgaria under Grant DCVP 02/1 (SuperCA++) and Grant DO 02-136-2008.

**Figure captions:**

**Figure 1**. Radial probability density of the Monte Carlo walkers for 3D para-helium (a) and for 3D ortho-helium (b). Blue lines – time-dependent Hartree-Fock (TDHF); red lines – TDQMC. The insets show the corresponding walker distributions in the x-y plane.

**Figure 2.** Time dependent ionization (a) and dipole moment (b) for 3D para-helium atom in an external electromagnetic field. Blue lines – time-dependent Hartree-Fock; red lines – TDQMC; green line – conventional Hartree-Fock. The inset shows the shape of the external electric field.

**Figure 3**. Time dependent ionization (a) and dipole moment (b) for 3D ortho-helium atom in an external electromagnetic field. Blue lines – time-dependent Hartree-Fock; red lines – TDQMC; green line – conventional Hartree-Fock.

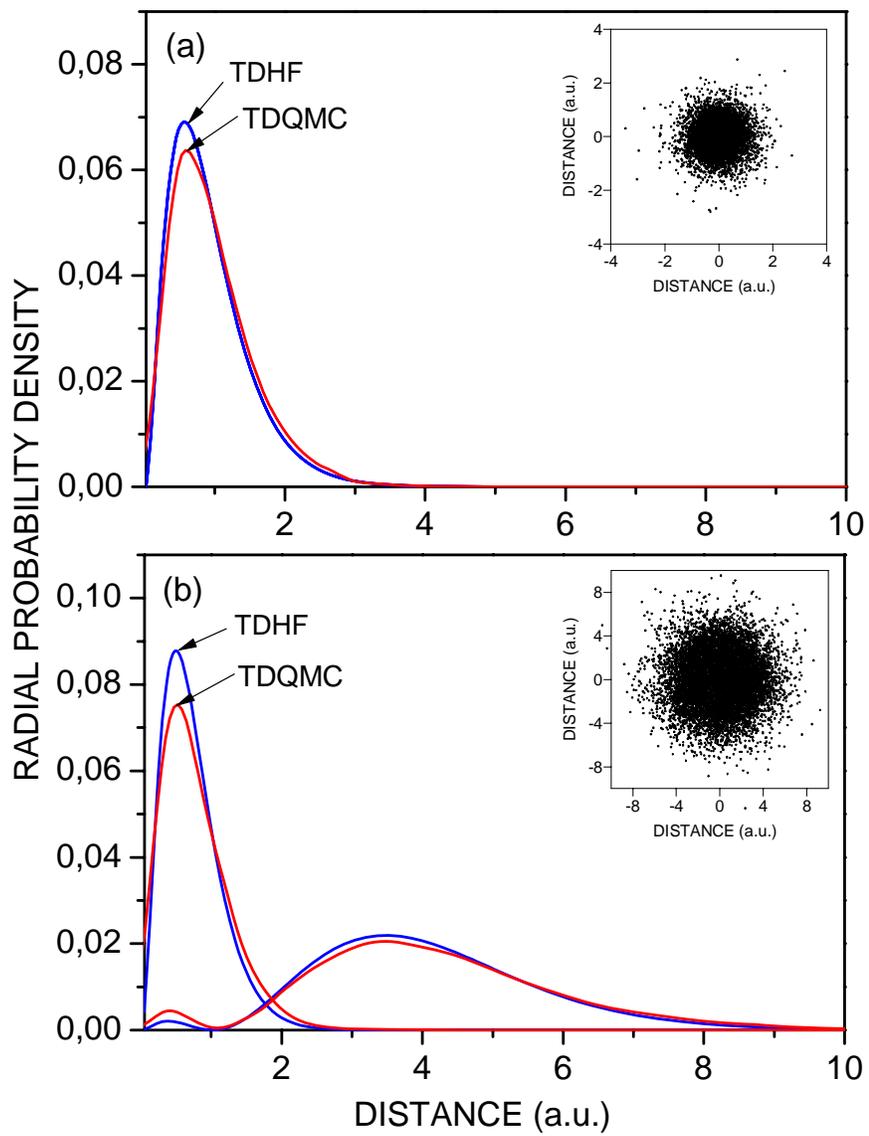

I. Christov, Figure 1

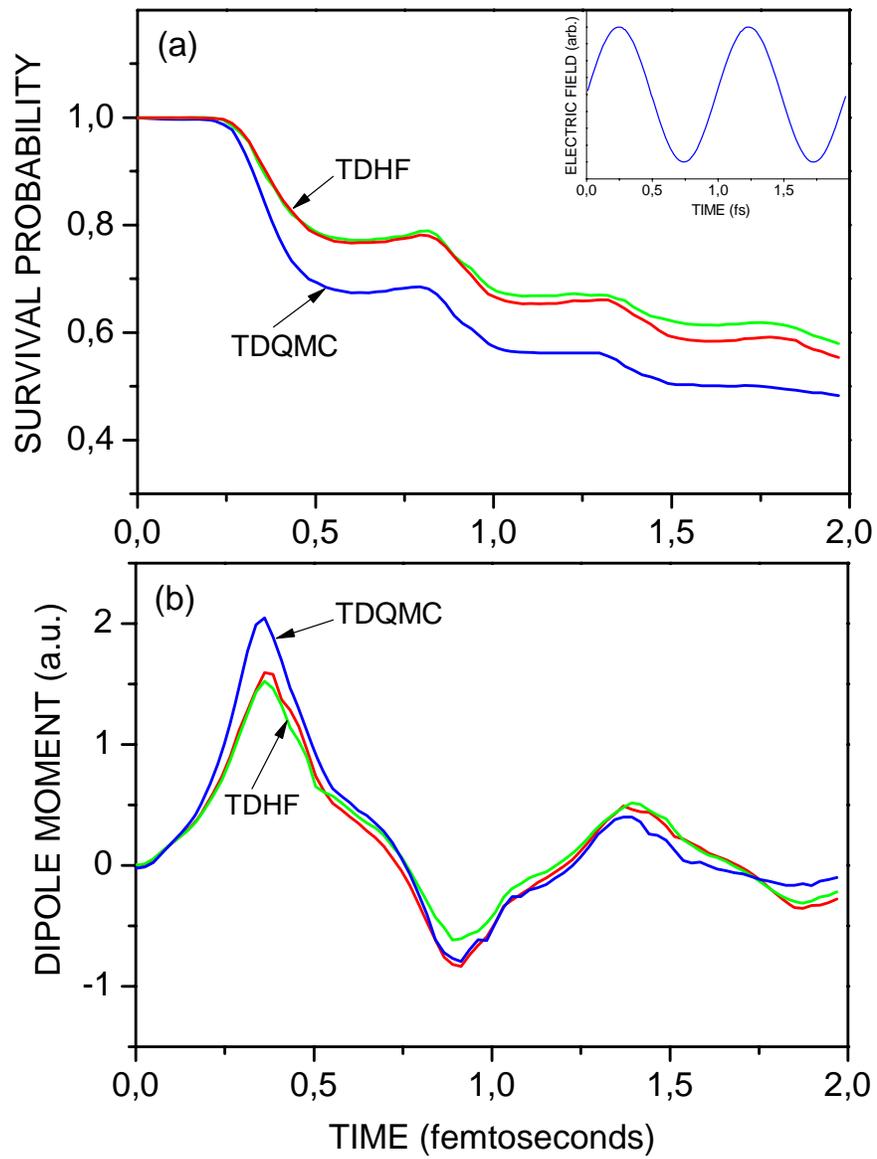

I. Christov, Figure 2

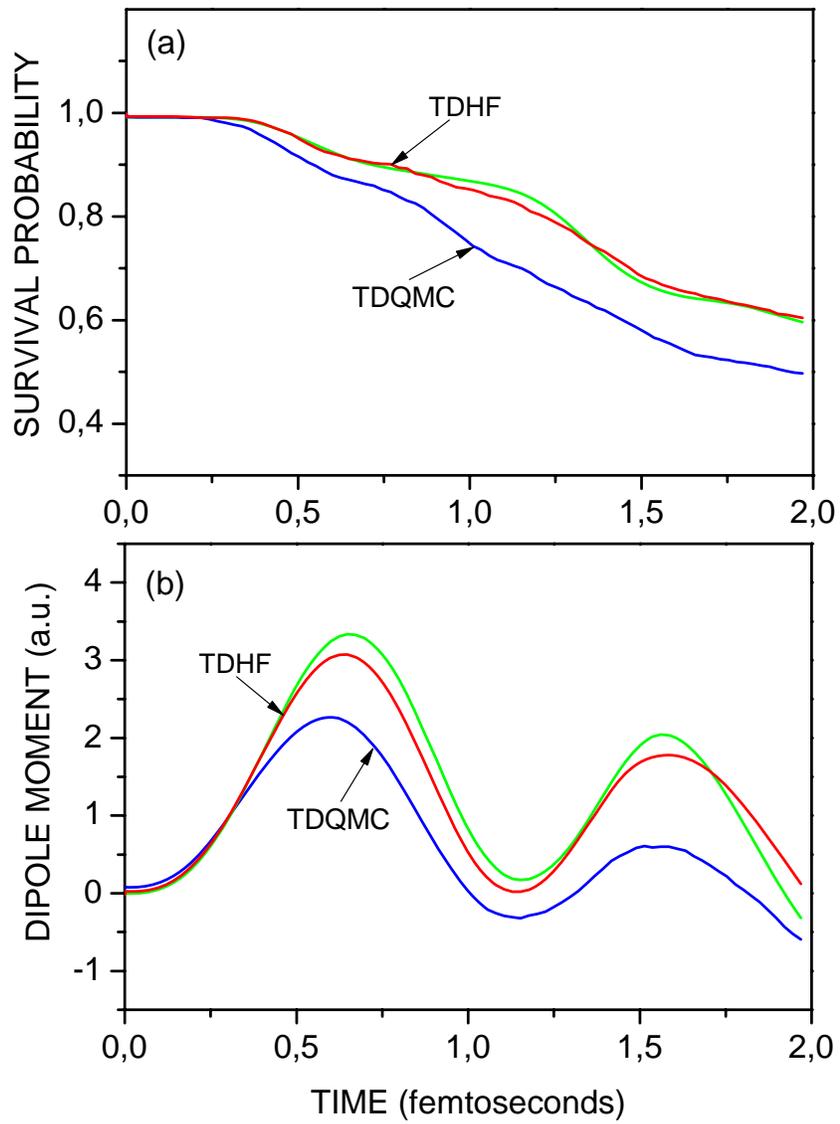

I. Christov, Figure 3

28